\documentclass[10pt]{iopart}
\usepackage{threeparttable}
\usepackage{graphicx}
\begin{document}

\title[JJ-Synapse]{Hybrid Synaptic Structure for Spiking Neural Network Realization}

\author{S Razmkhah$^1$, M A Karamuftuoglu$^1$ and A Bozbey$^2$}

\address{$^1$Department of Electrical and Computer Engineering, University of Southern California, Los Angeles, CA 90007 USA\\
$^2$Department of Electrical and Electronics Engineering, TOBB University of Economics and Technology, Ankara, Turkey}
\ead{razmkhah@usc.edu}
\vspace{10pt}
\begin{indented}
\item[]September 2023
\end{indented}

\begin{abstract}
Neural networks and neuromorphic computing play pivotal roles in deep learning and machine vision. Due to their dissipative nature and inherent limitations, traditional semiconductor-based circuits face challenges in realizing ultra-fast and low-power neural networks. However, the spiking behavior characteristic of single flux quantum (SFQ) circuits positions them as promising candidates for spiking neural networks (SNNs). Our previous work showcased a JJ-Soma design capable of operating at tens of gigahertz while consuming only a fraction of the power compared to traditional circuits, as documented in \cite{10109794}. This paper introduces a compact SFQ-based synapse design that applies positive and negative weighted inputs to the JJ-Soma. Using an RSFQ synapse empowers us to replicate the functionality of a biological neuron, a crucial step in realizing a complete SNN. The JJ-Synapse can operate at ultra-high frequencies, exhibits orders of magnitude lower power consumption than CMOS counterparts, and can be conveniently fabricated using commercial Nb processes.
Furthermore, the network's flexibility enables modifications by incorporating cryo-CMOS circuits for weight value adjustments. In our endeavor, we have successfully designed, fabricated, and partially tested the JJ-Synapse within our cryocooler system. Integration with the JJ-Soma further facilitates the realization of a high-speed inference SNN.
\end{abstract}
\vspace{2pc}
\maketitle
\ioptwocol
\section{Introduction}
Modern data management and processing demands the computational power to handle vast volumes of complex and interconnected information. Classical algorithms often prove inadequate for such tasks, necessitating the utilization of machine learning and deep learning algorithms for training computers to perform these operations effectively \cite{tavanaei_deep_2019,2903159,8259423}. While users typically employ conventional CPU and GPU architectures for deep learning and network training, the increasing reliance on this technology underscores the need for faster and more power-efficient computational resources \cite{wang2019benchmarking,Hager:20}.

Neural networks and neuromorphic architectures play pivotal roles in efficiently implementing learning algorithms in machines. By emulating the functionality of the biological brain, neuromorphic computers offer an alternative computational approach and a robust avenue for simulating Artificial Intelligence (AI) \cite{5275035,8052266}. Several neuromorphic architecture implementations, such as Intel's Loihi and Loihi2 chips designed for machine learning acceleration \cite{8259423}, rely on conventional CMOS logic. These designs integrate memory into the arithmetic logic unit, enabling in-memory computing and significantly enhancing the speed and robustness of the learning process. However, they continue to use voltage-level logic to mimic neural processes.

In contrast, research groups and companies, including IBM and Intel, explore innovative devices like memristors to replicate the learning and path formation observed in biological brain synapses \cite{Thomas_2013,8388703}. Memristors change their impedance as current flows through them, exhibiting behavior analogous to changes in ion channels and resistance drops akin to synapses \cite{6549211,6232461}. Nevertheless, memristors are associated with substantial power consumption, and their recovery time falls within the millisecond range, rendering them comparatively slow.

Superconductor electronics have long been regarded as a promising candidate for neural networks, with researchers exploring their potential for many years \cite{M_Hidaka_1991}. Single flux quantum (SFQ) logic, a key component of superconductor electronics, can operate at frequencies in the tens of gigahertz range while consuming power levels three orders of magnitude lower than traditional CMOS \cite{7383427,783712,5682046}. Moreover, the pulse-based logic inherent in SFQ is akin to biological neural signals, where the presence of a pulse represents logic one, and its absence denotes logic zero. SFQ cells are also inherently synchronous, relying on clock signals, with memory integrated into the logic. Although creating dense memory structures in SFQ can be challenging \cite{burnett2018superconducting,zha2022hiperrf,chen2020miniaturization,semenov2019very}, it is exceptionally well-suited for in-memory computation. The combination of neuromorphic architecture and SFQ technology is advantageous.

This work introduces a synapse structure called JJ-Synapse, leveraging a hybrid superconductor-semiconductor technology. Due to the quantized nature of SFQ pulses, the weights in the artificial JJ-Synapse are also quantized, necessitating training in a quantized manner. While this may slightly reduce algorithmic accuracy, it offers substantial advantages regarding on-chip memory and processing speed \cite{nagel2021white,yao2021hawq}.

Although SFQ technology offers remarkable capabilities, it is not self-sufficient for implementing a fully functional, high-performance neuromorphic processor. Resources such as biasing for individual neurons and storage for training parameters during execution are still required. To address this, we propose an SFQ-CMOS hybrid neuron cell that capitalizes on the strengths of each technology. CMOS components provide the necessary biases for JJ-Synapses and store weight values for the network in MOSFET switches. 
Si-Ge FETs are compatible with 4K operation, as shown by various groups \cite{Thomas2022}. In this work, we designed cryo-CMOS SG13G2 using the IHP 0.13 $\mu m$ BiCMOS process.
Since the CMOS only switches when configuring weights and remains fixed during inference operations, it does not impose significant speed limitations and has a negligible impact on power consumption.

We plan to integrate the proposed synapse design with our JJ-Soma cell, incorporating a quantizer/buffer circuit capable of handling negative and positive weights. This integrated unit, comprising JJ-Synapse, JJ-Soma, and a buffer, will emulate a neuron following an integrate-and-fire neuron (IFN) model. The resulting JJ-Neuron is expected to operate at nearly 20 GHz and maintain compatibility with SFQ logic, including RSFQ. This JJ-Neuron structure is a pivotal component for our future work, enabling the realization of a complete and fully functional neural network.

\section{Model and Methodology}
\begin{figure}[ht]
\centering
  \includegraphics[width=0.9\linewidth]{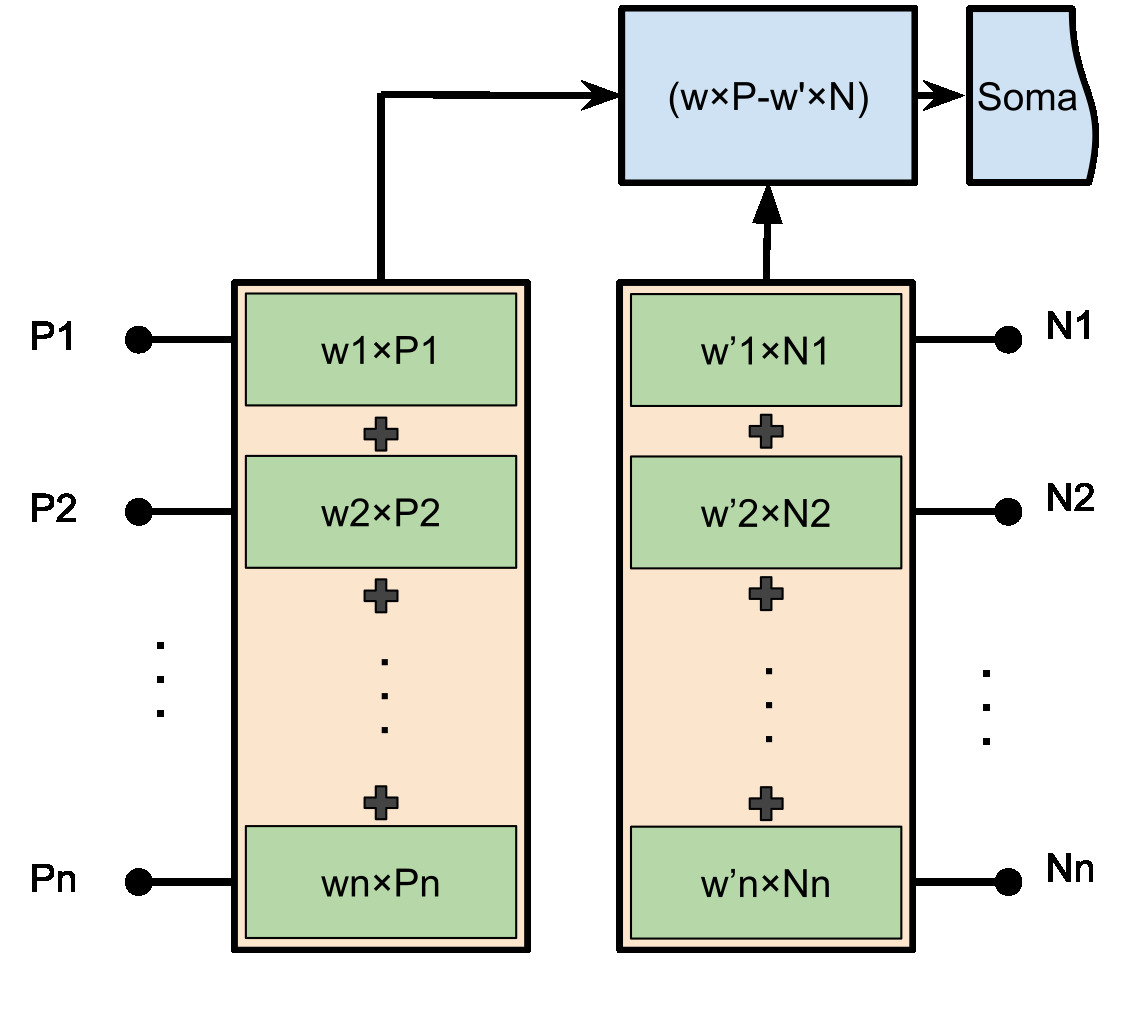}
  \caption{Block diagram of the JJ-Synapse with positive and negative inputs. Before applying this to JJ-Soma, inputs are summed in the Buffer/Quantizer (BQ) circuit.}
  \label{fig:1}
\end{figure}

A crucial aspect of implementing an artificial neural network is the presence of synapses with programmable weights that can accommodate diverse input levels, all while maintaining reasonable circuit size, speed, and power consumption. Moreover, these synapses should be versatile enough to process excitatory and inhibitory inputs. Given that our fundamental data unit is the SFQ pulse, the synapse must have the capability to generate a specific number of pulses, as determined by Equation \ref{eq:1}, at its output.
\begin{equation}
u = \sum_{k=1}^{n}(w_kP_k - w^{'}_kN_k)
\label{eq:1}
\end{equation}
In Equation \ref{eq:1}, the variables are defined as follows: $n$ represents the input port number, $w$ and $w^{'}$ are the weights, $N$ and $P$ can take on values of 0 or 1 based on the input value, and $u$ signifies the output of the JJ-Synapse, which will be applied to the JJ-Soma. When $u$ exceeds the predefined threshold, an SFQ pulse is generated at the output. Figure \ref{fig:1} illustrates this work's implemented model for the synapse structure. SFQ pulses are accumulated in both positive and negative branches, and then the two values are summed, with the output being applied to the soma. 

\begin{figure}[ht]
\centering
  \includegraphics[width=0.9\linewidth]{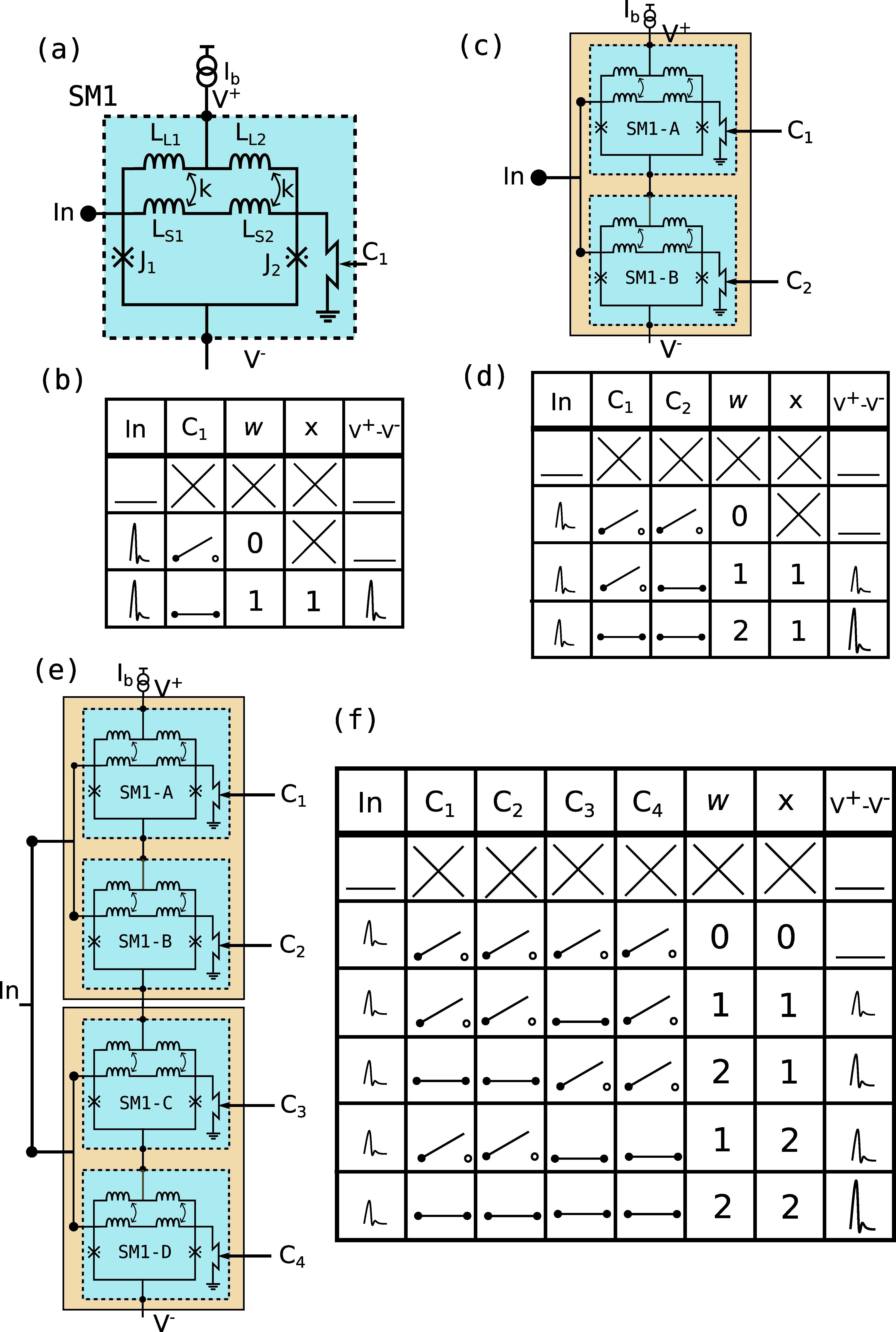}
  \caption{Design of the pulse accumulator. To accumulate pulses without losing any operating speed, we use a series of SQUID. (a) Demonstration of a single cell. (b) The truth table for a single cell as it switches. (c) Two connected cells make a weight of two for a single input. (d) The truth table for two cells. (e) Two connected cells of size two for two inputs that can weigh two. (f) The truth table for 2$\times$2 unit.}
  \label{fig:2}
\end{figure}

To align this model with SFQ and facilitate hardware integration, we assume that $P_k$ and $N_k$ equal one if an SFQ pulse is present and zero if no pulse is available. We also restrict $w_k$ and $w^{'}_{k}$ to integers between 0 and 4. As previously mentioned, imposing these limits may reduce network accuracy but can significantly enhance operating speed and simplify hardware integration.
Our synapse circuit comprises serially connected SQUID loops and input $LR$ circuits, each coupled with a SQUID loop and matched with the outputs of the SFQ library elements, as depicted in Figure \ref{fig:2}. The synapse circuit (SM1) 's building blocks are serially connected, significantly reducing bias current and power consumption. Each synapse circuit accepts spiking inputs and generates the positive or negative weighted output for the following layer. The switches $C_1 - C_4$ shown in Figure \ref{fig:2} are intended to be implemented using cryo-CMOS transistors.

Figure \ref{fig:2}(a) illustrates the circuit schematics of SM1. An applied SFQ pulse at the input (IN) propagates through $L_{S1}$ and $L_{S2}$, which are coupled to $L_{L1}$ and $L_{L2}$, respectively. This incoming pulse is coupled through these inductors, triggering the SQUID loop. Josephson junctions $J_1$ or $J_2$ switch and generate an SFQ pulse between nodes $V+$ and $V-$. Consequently, the circuit achieves the case where the weight $(w)=1$ and the input $(x)=1$. When the $C_1$ switch is open, current cannot flow from $L_{S1}$ and $L_{S2}$, resulting in a zero voltage difference on the result nodes. Thus, the circuit achieves a zero weight ($w=0$) in this configuration.

\begin{figure}[ht]
\centering
  \includegraphics[width=0.8\linewidth]{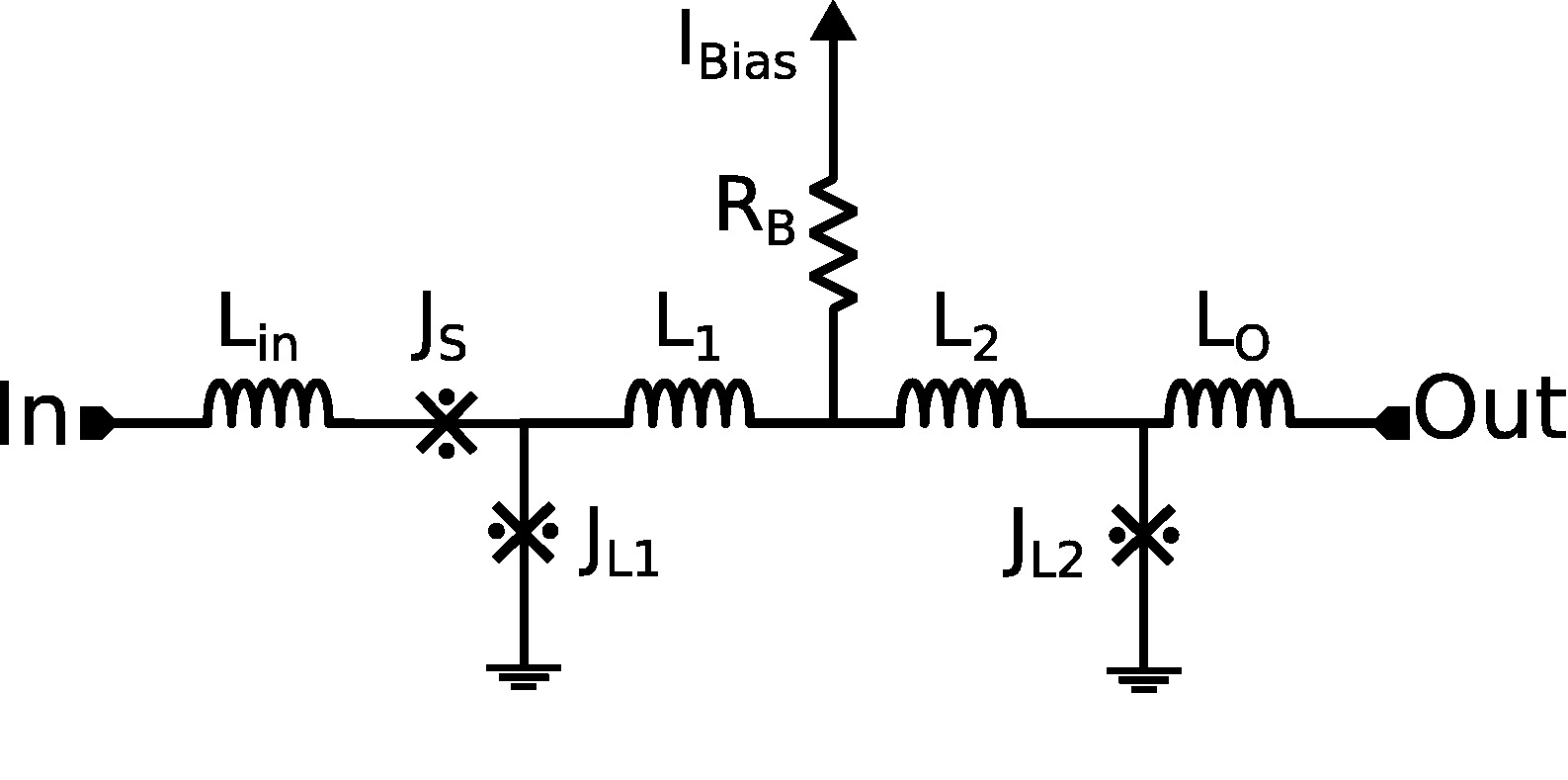}
  \caption{The Buffer/Quantizer (BQ) circuit functions similarly to the digital SQUID. BQ will sum the negative and positive fluxes applied to its loop and then provide quantized SFQ pulses based on the change in the flux. For symmetry, we have connected a negative load to the negative part of the circuit.}
  \label{fig:3}
\end{figure}

\begin{figure}[ht]
\centering
  \includegraphics[width=0.9\linewidth]{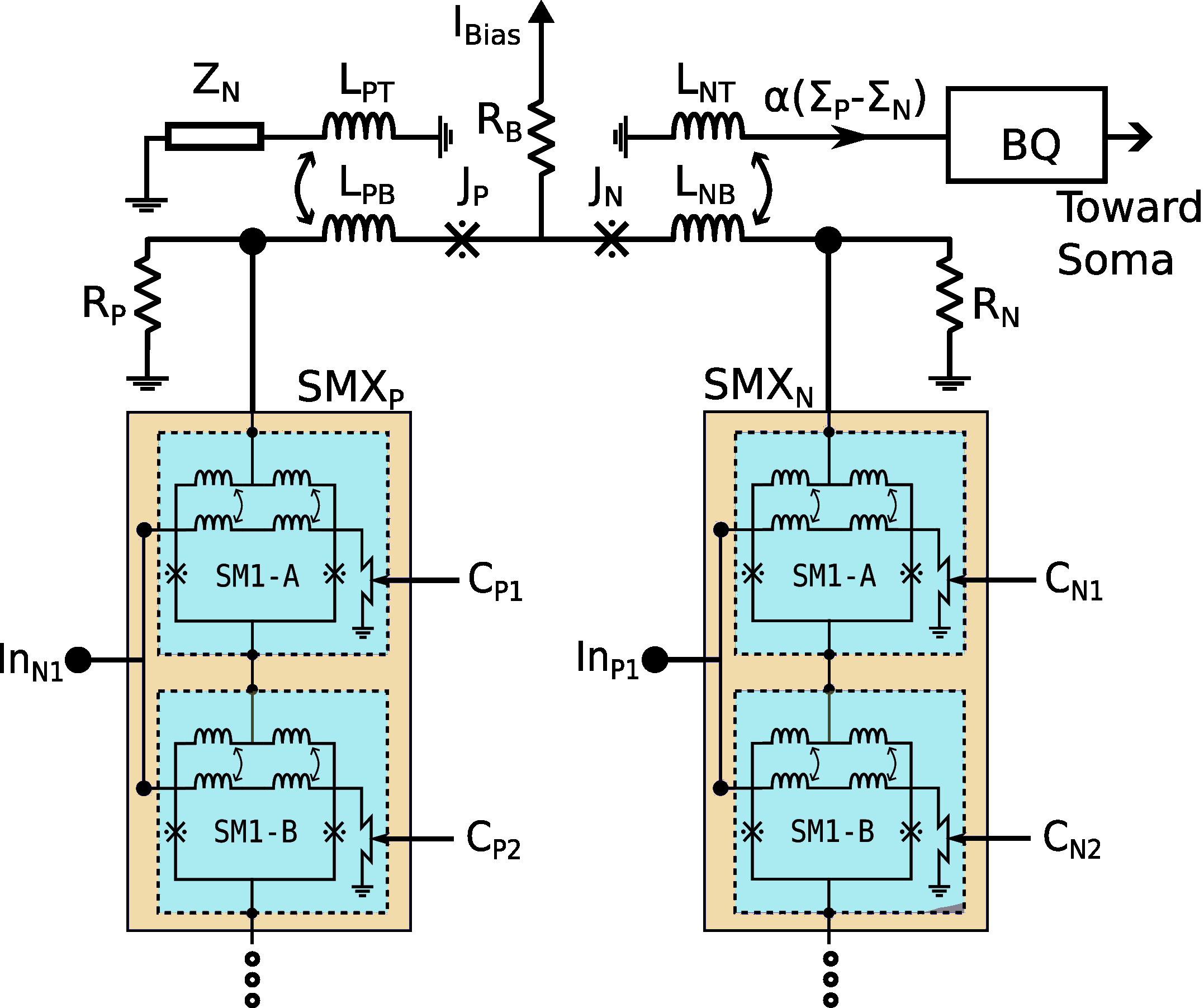}
  \caption{Schematic of the JJ-Synapse with positive and negative inputs. The Buffer/Quantizer (BQ) circuit has a negative load attached for symmetry.}
  \label{fig:4}
\end{figure}

We can generate weighted inputs by serially connecting these SQUID loops and providing the SFQ pulses simultaneously to each loop (SM2). For instance, by connecting two in series, we can implement weights 1 and 2 by controlling $C_1$ and $C_2$ based on the table shown in Figure \ref{fig:2}(d). When $C_1=1$ and $C_2=0$, the voltage difference at the output nodes equals one SFQ amplitude, resulting in a weight of one ($w=1$). Similarly, when $C_1=1$ and $C_2=1$, the difference equals two SFQ amplitudes ($w=2$). In addition to the weighted inputs, if the network requires input values ($x_i$) of 0, 1, and 2, four SM1 circuits can be serially connected to establish SM4. Based on the table shown in Figure \ref{fig:2}(f), it is possible to have $w_i=0$, 1, and 2 for $x_i=0$, 1, and 2.
A comparator circuit inspires the buffer/Quantizer (BQ) circuit, which acts similarly to an asynchronous quasi-one junction SQUID circuit. Figure \ref{fig:3} demonstrates the schematic of the BQ circuit. This circuit converts the cumulative energy of multiple pulses generated by SMX stages into several fast pulses proportional to $\sum P$, suitable for JJ-Soma input. Escape junctions $J_S$ are used to eliminate the SFQ backfiring of the circuit. We then connect the BQ circuit to the JJ-Soma for correct pulse application.
For a more accurate implementation of the neural network, it's essential that the synapses can handle negative weights. In Figure \ref{fig:4}, we've designed a circuit connecting two SMX circuits in a current-differential configuration. In this setup, the sum of positive weighted inputs ($\sum P$) and the sum of negative weighted inputs ($\sum N$) are differentiated to determine the net weighted inputs ($\sum P - \sum N$). Consequently, the current passing through the $L_{NT}$ inductor is directly proportional to the difference between $\sum P$ and $\sum N$ currents. The analog current value, $K$ times this difference ($K\times(\sum P - \sum N)$, where $K$ is the coupling factor), is converted into a quantized spiking configuration using the BQ circuit.

On the negative side of the circuit, a matching impedance, $Z_M$, is employed to match the impedance seen by the $L_{NT}$ and $L_{PT}$ inductors, ensuring that the positive and negative sides remain symmetrical and balanced. This configuration allows the synapse to handle both positive and negative weights effectively.
\section{Simulation Results}
\noindent
We conducted simulations of the circuits using the JSIM program, followed by an optimization process to enhance parameter margins. This optimization was carried out using a tool that employs the particle swarm optimization algorithm developed by our team \cite{tukel2013optimization}. The main objective of this optimization was to maximize the parameter margins of the circuit, making it robust against variations in the fabrication process. As a result of this optimization, we achieved parameter margins exceeding 20\%, which ensures the circuit's reliability even with manufacturing process variations. The bias margin, while relatively small at less than 5\%, is not a concern as it can be externally applied and finely adjusted for the circuit.

\begin{figure}[ht]
\centering
  \includegraphics[width=0.9\linewidth]{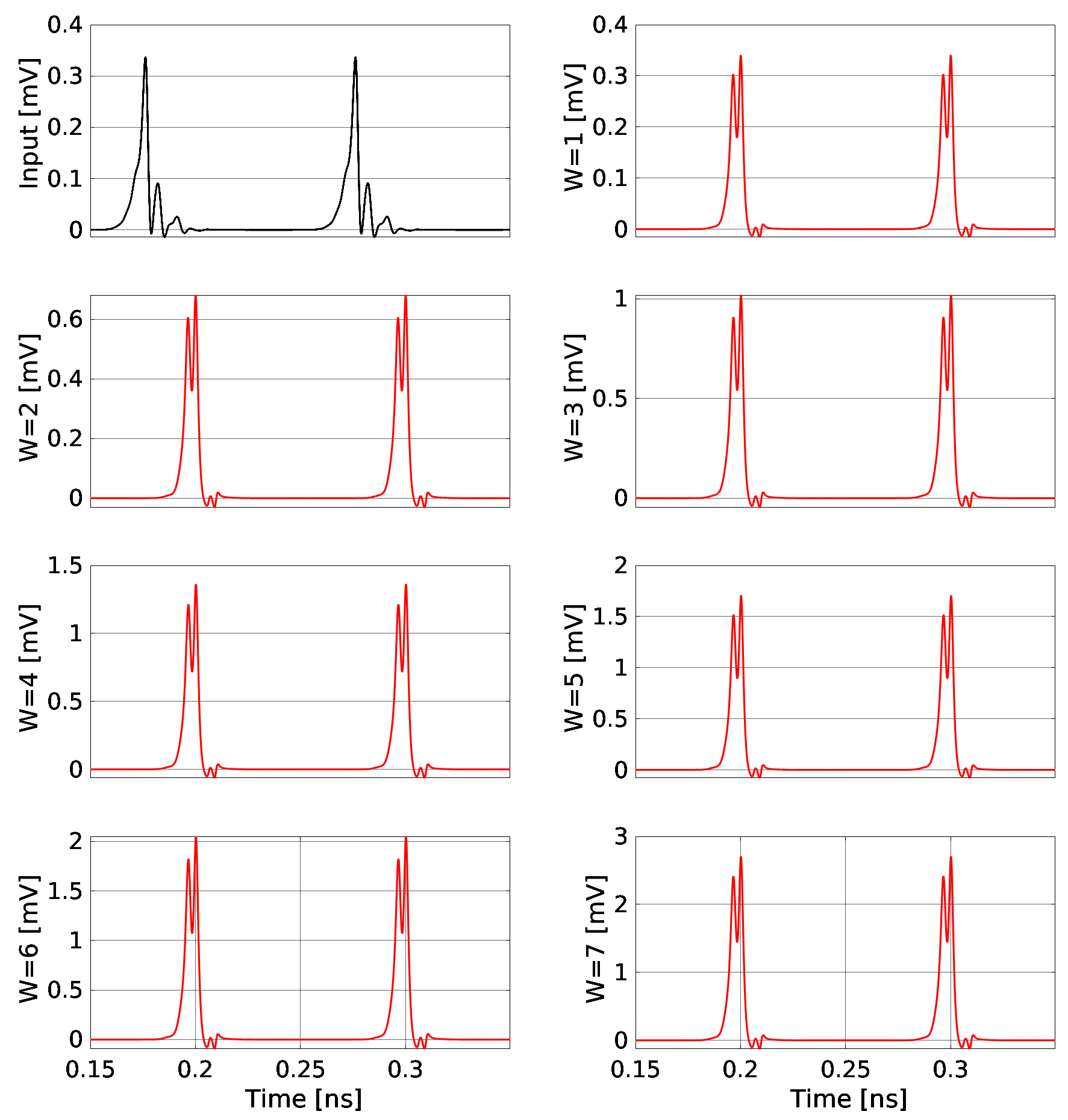}
  \caption{Simulation result of the JJ-Synapse with different weight values. The tested circuit is similar to the design shown in Figure.~\ref{fig:2}(e) but with 3-bit weight values. Here, we see the weight of the circuit changes as we switch the SQUID cells on and off}
  \label{fig:5}
\end{figure}

\begin{figure}[ht]
\centering
  \includegraphics[width=0.9\linewidth]{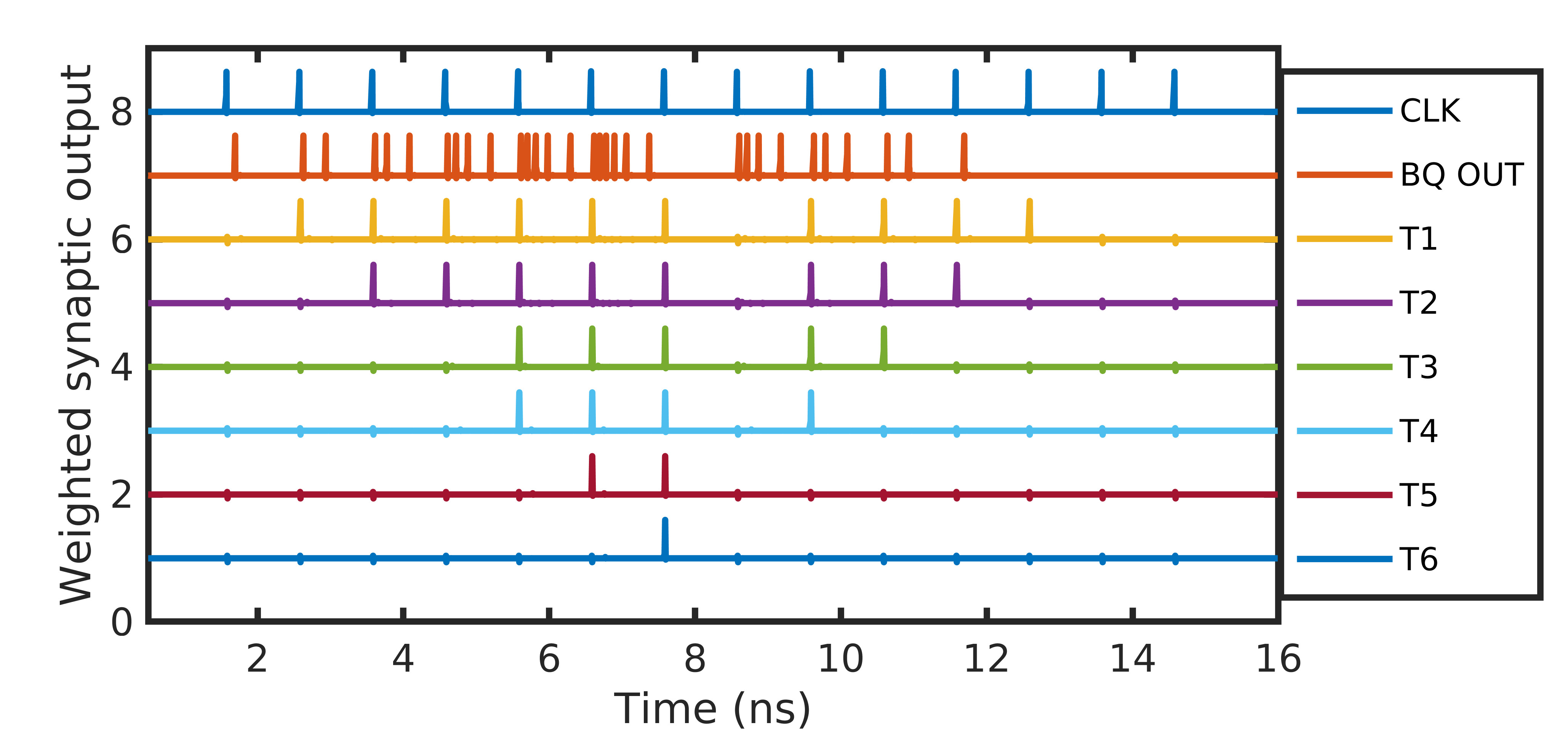}
  \caption{Simulation result for the circuit that is shown in figure \ref{fig:4}. The T1 to T6 are the thresholds of the Soma circuit.}
  \label{fig:7}
\end{figure}

We also simulated the JJ-Synapse structure depicted in Figure \ref{fig:2}(e). To enable 3-bit weight values, eight SQUID cells are required. These eight SQUIDs can be selectively turned on or off to achieve the desired weight value. Figure \ref{fig:5} presents the input signal and output of the JJ-Synapse for different weight values. It's worth noting that increasing the weight value results in higher signal amplitudes, increasing the energy. The applied input SFQ pulse had a frequency of 10 GHz, but it can be further increased to 25 GHz for a synapse with 3-bit adjustable weight capability.
We performed simulations of the circuit with various JJ-Soma thresholds and different weight values. The results of these simulations are presented in Figure \ref{fig:5}, which showcases the changes in circuit phases. Since each fluxon induces a 2$\pi$ phase change, the values are divided by 2$\pi$ for clarity. This figure measured negative and positive values from the JJ-Synapses' Josephson junctions and the Sum value obtained from the BQ circuit's output. Notably, if three or more fluxons are present at the BQ circuit output within a 50ps window, the soma will generate a pulse at the JJ-Soma's output.
\begin{figure}[ht]
\centering
  \includegraphics[width=0.9\linewidth]{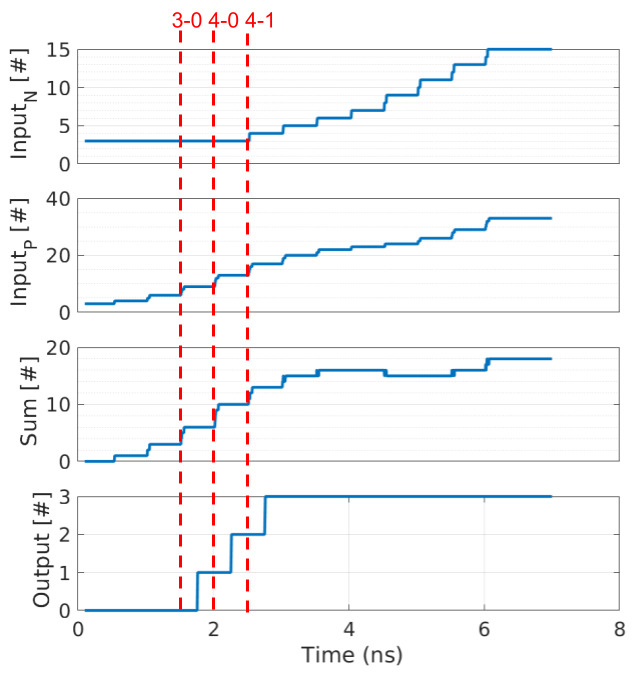}
  \caption{Simulation result of the neuron circuit with threshold 3. Here, we measure the phase and divide it by 2$\pi$ to get the number of pulses. The negative and positive inputs are measured from the synapses, and the Sum is the output of the BQ circuit. The output pulse is generated when three fluxes are in the BQ circuit in less than 50ps.}
  \label{fig:6}
\end{figure}
To demonstrate the circuit's functionality, we designed a 4-input synapse with weight values of {0, 1, 2, 3} in both positive and negative parts, effectively supporting 3-bit weights. We then connected this circuit to a JJ-Soma with threshold values ranging from 1 to 6. The resulting outputs are illustrated in Figure \ref{fig:6}. It's important to note that these circuits can be further stacked to accommodate 4-bit weight values and 4-bit threshold values, making them suitable for mapping most neural networks onto the presented structure. Figure \ref{fig:7} demonstrates the simulation results for when different synaptic weights from 1 to 6 are applied to the neurons with threshold values ranging from 1 to 6.  
\section{Hardware Implementation}
\begin{figure}[ht]
\centering
  \includegraphics[width=0.6\linewidth, angle=90]{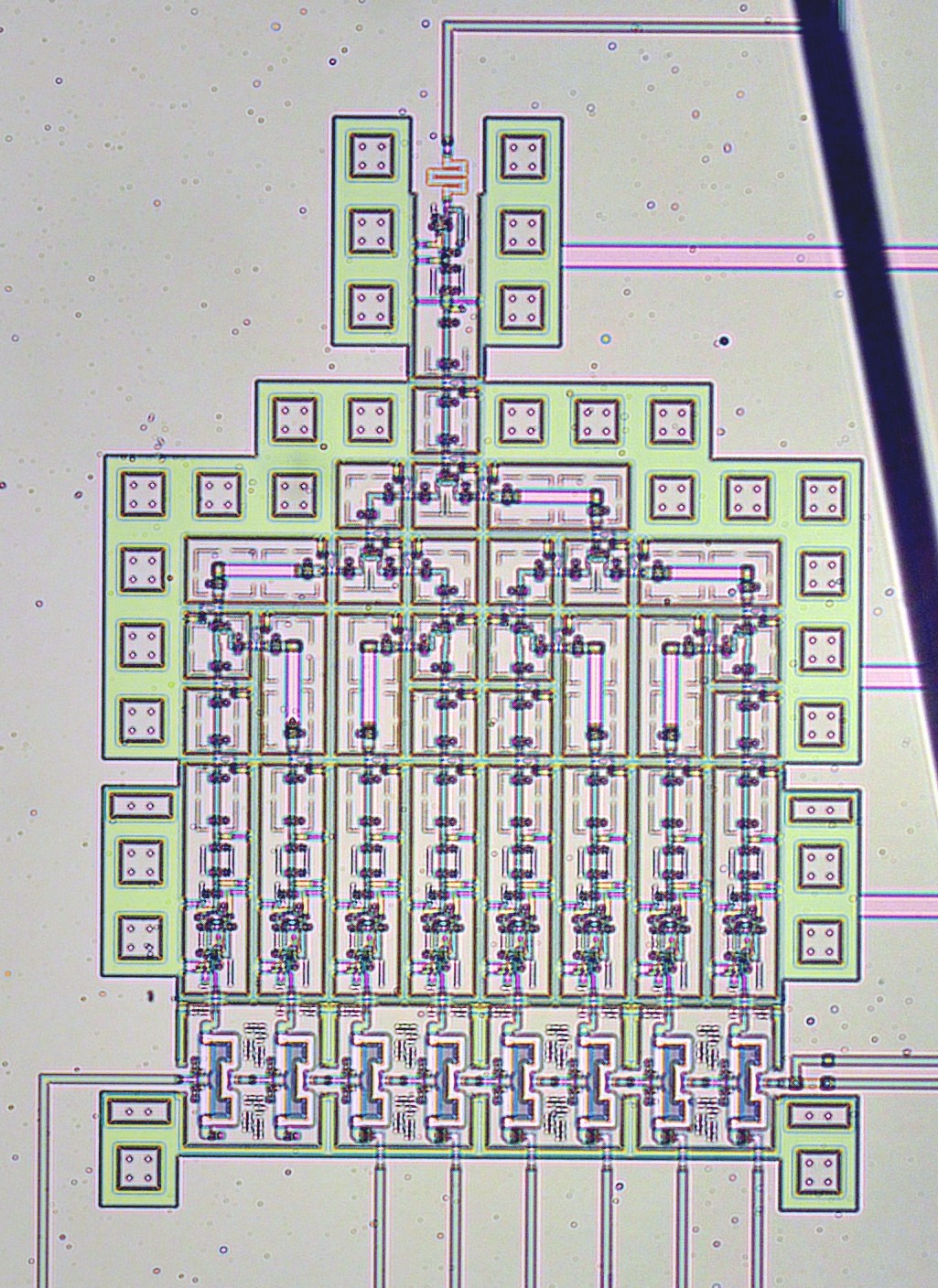}
  \caption{Test circuit layout designed with HSTP process to confirm the functionality of JJ-Synapse. The wires from SQUIDs are connected to the pads that will be wire-bonded to a Si-Ge CMOS chip for switching.}
  \label{fig:8}
\end{figure}

\begin{figure}[ht]
\centering
  \includegraphics[width=0.8\linewidth]{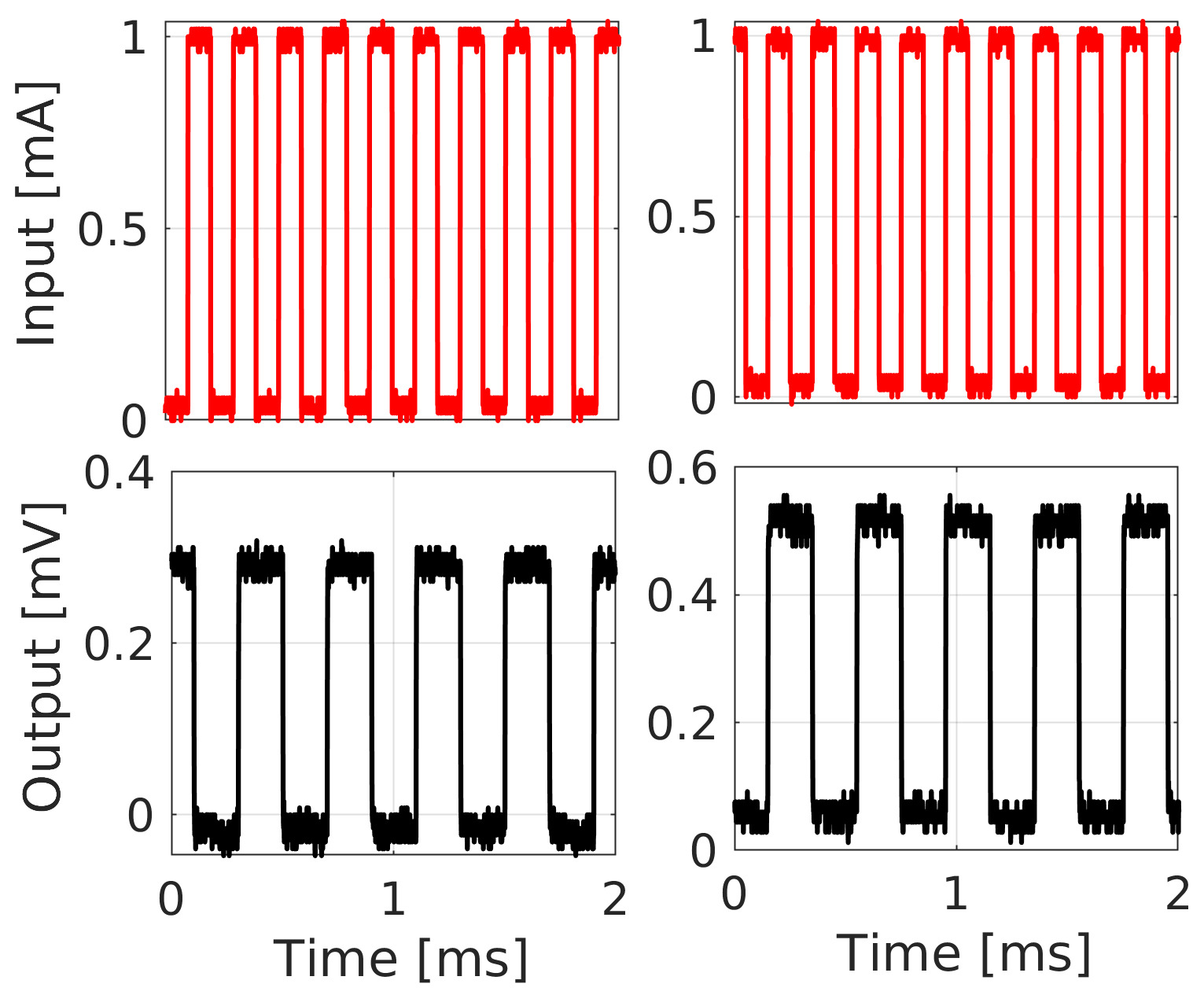}
  \caption{Measurement result for weights 2 and 4 on the left and right, respectively. Here, weight four is not scaled linearly and is 1.5$\times$ of weight 2.}
  \label{fig:9}
\end{figure}

\begin{figure}[ht]
\centering
  \includegraphics[width=0.9\linewidth]{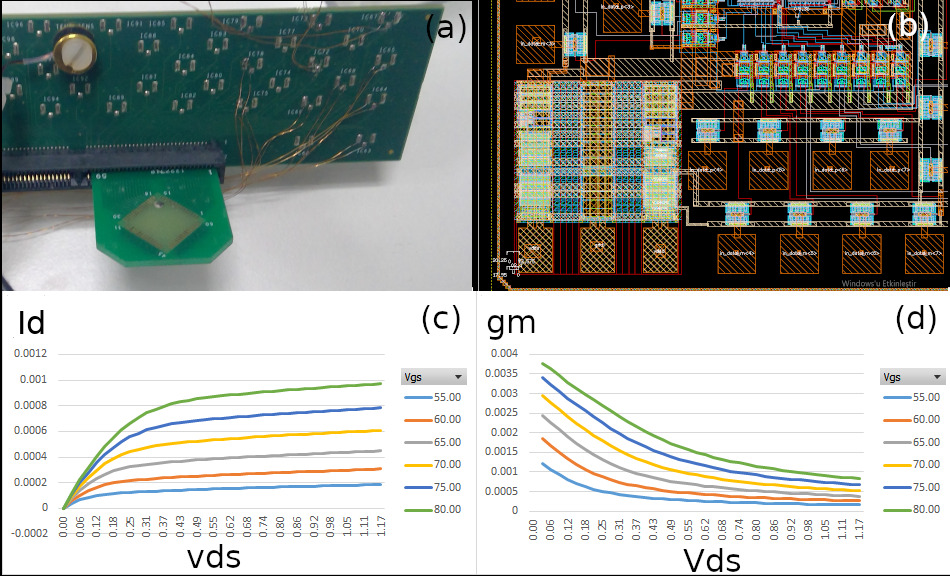}
  \caption{The SiGe cryogenic CMOS is designed for implementing the synaptic weights. (a) demonstrates the chip on the test probe that was lowered in the cryostat with the temperature sensor at the top left corner of the probe, (b) is the layout of the chip designed for IHP SiGe process, (c) is the $I_d$ vs. $V_ds$ graph at 4.2K in different $V_gs$, and (d) is the $g_m$ vs. $V_ds$ measurements at 4.2K in different $V_gs$ values.}
  \label{fig:10}
\end{figure}

We designed the test bench layouts for a sample synapse circuit to investigate its operation. AIST CRAVITY fabricated the circuits with a 4-layer Nb process HSTP \cite{hidaka2006current}. Figure \ref{fig:8} shows the fabricated test bench. The circuit has a DC/SFQ converter cell as input, and the number of SFQ pulses increases by a splitter tree. Here, the default weight is two and will increase as the other SQUIDs connect. A bi-CMOS-based cryo-circuit will switch the SQUID connections on or off during the operation of the final circuit. However, in the tests, no CMOS was used. The experimental setup for this circuit is presented in \cite{razmkhah2021compact,razmkhah2019setup}.
The measurement results for a default weight of 2 and an increased weight of 4 are demonstrated in Figure \ref{fig:9}. Since we cannot capture an SFQ pulse, we implement the amplifier after an SFQ/DC converter circuit stage. The output that we measure on the oscilloscope is the average value of all the combined SFQ/DC pulses via the SQUID stack, and therefore, the output of 4 stages is not scaled linearly but is about 1.5$\times$ of weight 2.
To confirm the functionality of CMOS switches, we designed and fabricated a SiGe-based CMOS chip and measured it in a gas transfer type GM cryocooler. The test probe, circuit layout, and measurement results of an NMOS at 4.2K are shown in Figure \ref{fig:10}. The $I_d$ vs. $V_ds$ graph confirms the operation of the NMOS at 4.2K. Since the CMOS will not switch during the operation of the SFQ chip, its speed and heat generation are negligible, and it can work at sub-threshold to minimize its effect on SFQ circuits. The CMOS chip will sit at the top of the SFQ chip and will be bonded to it.
\section{Conclusion}
\label{sec:conc}
\noindent
We have designed and fabricated a neuron circuit using superconductor circuits. We will later use the proposed design to realize a spiking neural network. The JJ-Synapse uses a series SQUID circuit with no static power consumption. The power is consumed in SFQ pulse switching and interconnects between the cells. The circuit is compact, scalable, and has an acceptable bias margin value. The resulting SNN will be programmable; users can modify it in real-time. The neuron's weight and threshold are variable via an in-situ Si-Ge CMOS circuit. We fabricated the JJ-Synapse, and experimental measurement done on the pulse multiplication confirmed the circuit's functionality. The JJ-Neuron can function up to $20 GHz$ and have power consumption in the order of attojoule for a single operation. We simulated the whole circuit with JSIM and will test the circuit with the CMOS chip as a controller, and later, we will demonstrate the complete SNN network.
\ack
\noindent
The circuits are fabricated in the clean room for analog-digital superconductivity, CRAVITY (currently known as QuFab) of the National Institute of Advanced Industrial Science and Technology (AIST).

\end{document}